# Relativistic Accretors and High Energy X-Ray view

Asish Jyoti Boruah*; Liza Devi; Biplob Sarkar
* Corresponding author
Department of Applied Sciences, Tezpur University, Napaam, Assam-784028, India
E-mail of corresponding author: app23109@tezu.ac.in

*Abstract*—Relativistic accretors are cosmic objects that pull matter from their surroundings at speeds almost equal to the light's speed. Because of the tremendous gravitational force from the accretors and the angular momentum of infalling material, which often result in discs of gas and dust that are heated to extremely high temperatures. We encounter strong radiation throughout the electromagnetic spectrum, including intense X-rays. The X-ray view provides a unique window into the behavior of accretors. In this review, we discuss different accretors, particularly binaries, and their origin, involved mechanisms, and properties, including energy spectra and the variability of X-rays from the accretors. This X-ray perspective gives a unique insight into the evolution and connections of these systems with their environment. Future research in this area is necessary to fully understand the process underlying X-ray emission from relativistic accretors.

## I. INTRODUCTION

Relativistic accretors are astronomical objects that grasps nearby material through their enormous gravitational fields, which leads to the phenomenon that challenges our understanding of physics. These systems are generally associated with compact objects (COs), notably as matter reaches the speed of light via the process of accretion. When matter spirals towards these objects, we observe a formation of accretion disc (AD), which is actually the hot, whirling mass of dust and gases. The heat generation occurs due to friction and gravitational forces, and emanate electromagnetic radiation like X-rays and gamma rays. Due to this, they become the most luminous objects in the cosmos. Within the framework of present-day astrophysics, relativistic accretors are essential to a number of cosmic occurrences, such as the dynamics of binary star systems and the production of relativistic jets [1].

As previously stated, accretors emit high-energy radiations in a variety of electromagnetic wavelengths. However, our study in this article will focus solely on the X-ray range. Typically, X-rays have wavelengths between 0.01-10nm. X-rays are primarily divided into two categories based on their wavelength: Soft X-rays (SXR) (0.1-10 keV) and Hard X-rays (HXR) (10-100 keV). Wilhelm C. Roentgen first discovered X-rays in the last decade of 19[th] century during the study of electric discharge phenomena in gas. After that significant finding, X-rays from astronomical sources were not identified until the improvement of rocket technology after 2[nd] World War. The first report of X-rays from powerful cosmic source (Sco X-1) was published by Riccardo Giacconi [2], which was the initial point for scientific community to start research in astrophysical X-ray sources. The role of X-ray observations is very crucial in accretion physics because they offer significant insights into the physical mechanisms that regulate accretion onto accretors. X-ray observations offer diagnostics of the AD, such as X-ray spectra and variability. This reveal disc temperature, structure, and density. Jet dynamics and emission can be followed via X-ray observations. It also investigates the evolution of binary systems and companion star features.

## II. TYPES AND PROPERTIES OF ACCRETORS

Uncovering the mysteries of some of the most fascinating and intense astrophysical phenomena requires an understanding of different kinds of relativistic accretors such as black holes (BH), neutron stars (NS), white dwarfs (WD), and binary system and their characteristics. These are briefly discussed below:

### A. Black Hole

These are the most extreme types of accretors, possessing an event horizon [3]. A tremendous amount of energy is released by gravitational interactions when materials compress and acquires kinetic energy as it spirals towards a BH, resulting in an AD. Prominent example of BH accretors include Sagittarius A*, BHs in binaries like Cygnus X-1 and GRS 1915+105 [4]. Mall et al. [5] reported that they analysed one accreting BH known as "XTE J1859+226" to measure the spin using X-ray reflection spectroscopy. They discovered a contradiction with the result using the relativistic precession model. Studying BH transient Swift J1727.8-1613., Veledina et al. [6] reported the detection of X-ray polarisation for the first time using Imaging X-ray Polarimetry Explorer (IXPE).

### B. Neutron star

NS can be observed at the end stage of massive star's evolution ($M > 8M_\odot$). NSs are the collapsed cores of massive stars which as a result undergoes a supernova (SNe) explosion and have powerful magnetic fields (MFs) and comprise mainly of neutrons [7]. There are two primary ways through which accretion onto NSs might happen. First, Roche lobe overflow from a companion and secondly, through stellar wind accretion. NS accretors are known as X-ray pulsars, which emit pulsed X-rays caused by the tilt of the NS's MF. Examples include the Crab Nebulae pulsar and the Vela pulsar [8]. Using the whole set of data gathered by the Insight-HXMT and the Fermi/GBM outcomes, Tuo et al. [9] examined the temporal evolution of the coherent X-ray pulsations observed by 4U 1901+03 during its 2019 outburst. They discovered that during the outburst, the pulse profiles drastically changed.

### C. White dwarf star

The remnants of stars that have exhausted their nuclear fuel and have collapsed due to their own gravity are recognized as WD. They have a relatively small size and mass but high surface temperatures. WDs have entirely "pressure ionised" materials, comprise of degenerate electron gas [10]. Accretion onto WDs occurs through overflow of Roche lobe from a companion star. WDs are frequently observed in novae and cataclysmic variable (CV) stars [11]. XMM J152737.4−205305.9, a polar-type CV recently found with the XMM-Newton, was discovered and further investigated, as reported by Ok et al. [12]. This newly discovered CV has deep eclipse-like features.

*D. X-ray binary (XB) system*

These systems are comprised of a NS or BH accreting materials from a donor (usually a typical main sequence star). XB can be categorized into high mass XB (HMXB) and low mass XB (LMXB).

- HMXB: In HMXB, one of the components is a NS or BH, and the other is a massive O or B type star ($M \geq 5M_\odot$) [13, 14]. HMXBs normally originate when a massive star undergoes nuclear fusion, gradually exhausting its nuclear fuel. As it expands, it loses its outer layers in the form of a strong stellar wind. The CO gravitationally grasps some of this material, resulting in an AD that is transient in nature [13, 15]. In our Milky Way galaxy, there is an unusual HMXB known as Cygnus X-3 that has been observed often, particularly by INTEGRAL, as it is one of the seven known Wolf-Rayet XB [16].
- LMXB: In LMXB, there is also a NS or BH on one side and the other one is the donor star, but here accretion of materials by CO occurs through Roche lobe filling and disc formation. In LMXB, donor star's mass is less compared to HMXB's donor star (i.e. $M \leq 1M_\odot$) [14]. The spectroscopy given by six simultaneous observations from XMM-Newton and Rossi X-ray timing Explorer (RXTE) of the NS LMXB 4U 1636–53 suggests that there could be a relationship between the height of the corona and the normalization of the disc emission [17].

*E. Jet-Producing Systems*

Some relativistic accretors, especially BHs, generate strong relativistic jets. Jets are energetic beams of ionized gas that are collimated and released near the accretor, propelled by rotational energy extracted from the AD. These systems have the ability to interact with the surrounding interstellar medium and span large distances. Few examples of these systems include active galactic nuclei (AGN) and micro-quasars [15]. The first direct observational constraints on the kinematics of the jet-producing zone in a BH jetted system was reported by Punsly [18]. He stated that recently captured 86 GHz image of M87 reveals a jet emerging from the AD around a BH.

### III. MECHANISMS OF X-RAY EMISSION

Astrophysical X-rays from relativistic accretors are predominantly generated by the following three processes:

*A. Thermal bremsstrahlung (TB)*

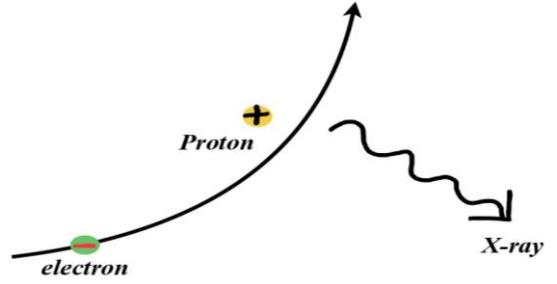

**Fig. 1. Schematic Diagram of Thermal bremsstrahlung.**

When atoms of hot, thin, and transparent gas are ionised at temperatures exceeding $10^5$ K, thermal energy is transferred between the particles via collisions. When an electron approaches a positive ion, a shift in its trajectory is detected due to significant electric force, which causes the electron to accelerate and emit X-rays. This radiation is known as braking radiation, or bremsstrahlung. This radiation is continuum in nature having a distinctive shape based only on temperature, and it is called TB [19]. For example, after analysing data from the RXTE for the BH candidate SS 433, Seifinai and Titarchuk [20] discovered that the TB spectrum with a temperature of 10–30 keV approximates the broad band (BB) continuum (up to 100 keV) spectra. Wilkins et al. [21] jointly analyzed the flare data from Swift and NuSTAR observations of the supermassive BH in Mrk 335. During the spectrum analysis, they observed the existence of TB, photoelectric absorption, and Compton scattering.

*B. Synchrotron radiation (SR)*

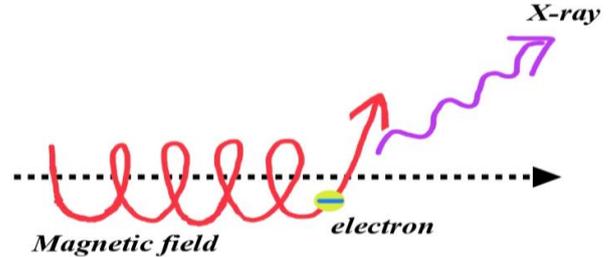

**Fig. 2. Schematic Diagram of Synchrotron radiation.**

In the vicinity of MF, an electron travelling fast, having a range of velocities experiences an electromagnetic Lorentz force. Due to the change in velocity vector, the relativistic electron is accelerated, resulting in the emission of X-rays. This process is termed as SR or magnetic bremsstrahlung. SR frequency is entirely determined by the energy of the electron, the strength of the field, and the direction of electron's motion [19]. For example, X-ray emissions from GRS 1915+105 show variability that is consistent with synchrotron radiation [22]. Jin et al. [23] reported XMM-Newton observation of a highly super Eddington narrow-line Type-1 quasi-stellar object (QSO) RX J0439.6−5311. The report includes SR along with SXR excess and is dominated by a low temperature, optically thick Comptonization component rather than relativistic reflection.

## C. Inverse Compton scattering (ICS)

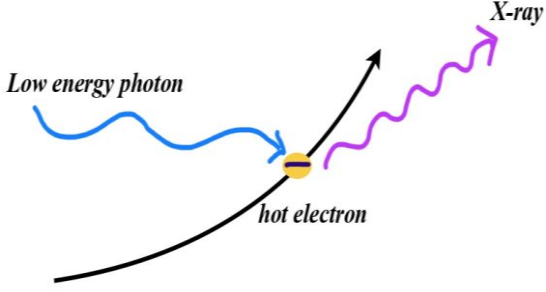

**Fig. 3. Schematic Diagram of Inverse Compton scattering.**

It occurs when electrons moving at extreme relativistic speed scatter low energy photons to high energies, allowing the photons to acquire energy at the expense of the kinetic energy of the electrons, resulting in high energy X-rays [24]. Wilkins and Gallo [25] takes into consideration the Comptonization of the photons constituting the relativistically blurred reflection typically observed from AGN ADs. They believed that the coronae of energetic particles produce the strong X-ray continua through ICS of thermal seed photons from the disc. Adegoke et al. [26] reported the first direct observational confirmation through the study of Mrk 493 that the AD is the source of the seed photons required for thermal Comptonization in the hot corona, consequently constraining the corona's size to 20 Schwarzschild radii. The HXR spectrum of the AGN for the bright X-ray Seyfert 1.2 galaxy IC 4329A is generated by the corona's Compton up-scattering of the optically thick disc photons [27].

## IV. X-RAY VIEW OF ACCRETORS

X-ray emission from relativistic accretors ranges in energy from SXRs coming from the AD's innermost regions to HXRs generated by highly energetic processes such as Compton scattering in hot, dense electron cloud (coronae) or magnetic bremsstrahlung from relativistic jets [1]. Based on accretion state, relativistic X-ray sources can have significantly distinct X-ray spectra:

- High-Soft State (HSS): With strong emission lines and a relatively SXR spectra, this state is illustrated by a dominating thermal component originating from the AD. In general, this state is associated with a stable inner AD and higher accretion rates [28].
- Low-Hard State (LHS): This state is characterized by low or absent thermal emission lines and a power-law driven spectrum. A power law-driven spectrum is a form of frequency spectrum in which the signal's amplitude varies with frequency according to a power law distribution [1]. Stronger (higher energy) X-ray emission is frequently linked to a truncated inner disc, low accretion rates, and substantial jet activity [28].
- Intermediate States: As the accretion rate varies dynamically, transitions between high/soft and low/hard states may yield complex X-ray spectra, including collections of thermal and non-thermal components [28]. Based on variability, two other states are distinguished from this one: Hard intermediate state (HIS) and Soft intermediate state (SIS). BB fractional root mean-squared variability is less in the HIS than in the hard state. A weak power-law noise component, which takes the place of the BB noise component found in the HIS, explains SIS [29].

Table I. Table for summarizing states of different X-ray spectra.

| State | Accretion rate ($\dot{M}$) | X-ray spectrum | Photon index ($\Gamma$) | Reference |
|---|---|---|---|---|
| HSS | High ($\dot{M} = \dot{M}_{Edd}$ or higher) | Dominated by thermal disc emission | $\Gamma > 2$ | [28] |
| LHS | Low ($\dot{M} < \dot{M}_{Edd}$) | Dominated by non-thermal emission | $1.5 < \Gamma < 2$ | [28, 30] |
| Intermediate | Moderate $\dot{M}$ | Mixed spectrum of both thermal and non-thermal emission | $\Gamma \sim 2$ | [28] |

### A. Energy spectra

There are various models that have been proposed by groups of scientists when it comes to the study of X-ray emission. Among all those models, the truncated disc model has been demonstrated to be consistent with X-ray spectra, rapid X-ray variability, light curves and relativistic jet behavior. This model is consistent with both BH and NS. In this model, the zone between the inner edge of the disc and the innermost stable orbit of accretor is replaced by a hot, optically thin, geometrically thick accretion flow (AF) [28]. Together with it coexists an optically thick, slim Shakura-Sunyaev (SS) disc that is truncated at specific radius at low luminosities. At this point, very few photons from the disc illuminate the flow. In comparison to the heating resulting from protons' collisions, the electrons' Compton cooling is nearly inefficient. The shape of a thermally Comptonized energy spectrum is primarily determined by two parameters: the optical depth of the plasma and the power ratio (i.e. the luminosity ratio between hard state and soft state). In the hard state, hard thermal spectra are generated due to a small number of seed photons creating the hot inner flow, described as a power law, $P_E(E) \propto E^{-\Gamma}$ with a photon index of $1.5 < \Gamma < 2$ in the 5–20 keV energy range [1, 28]. The inward motion of the disc causes it to extend under the hot inner flow, which raises the number of seed photons and lowers the power ratio mentioned above. Consequently, softer spectra and higher frequencies in the power spectra result from a decrease in the truncation radius. Spectra resulting from this have a combination of soft and hard spectral components that are observed during the short-lived SIS and HIS. Significant spectral and temporal changes are observed when the truncation radius approaches the innermost stable orbit. At this point hot flow may collapse into an SS disc [28]. The existence of the inner disc causes a large rise in disc flux, which indicates the hard-soft-state transition. Additionally, it implies that any residual electrons that obtain energy outside of the optically dense disc material will experience a stronger Compton cooling, leading to softer spectra. A power law with $\Gamma \geq 2$ and energy exceeding 500 keV generally describes the soft state, which is characterised by a powerful disc and soft tail [31]. Unlike the hard tail observed in the LHS, this tail is not the result of thermal Comptonization. The spectrum should be generated in a zone with a high temperature and relatively shallow optical depth if it is to extend to 500 keV and beyond [28].

*B. Rapid temporal variability*

Power density spectra (PDS) are typically examined in order to study rapid variability in XBs. The majority of the broad power spectrum components in PDS of BH Binaries may manifest as either a more localized peak (quasi-periodic oscillations, QPOs) or as a broad power distribution spanning many decades of frequency [32]. QPOs are a typical feature of accreting BHs. They are also detected in binaries of NSs, in CV, AGNs. Primarily there are two categories of QPOs based on frequency.

- Low-Frequency QPO (LFQPO): They have frequencies between a few mHz to roughly 30 Hz. A variety of QPO types are presently recognised based on their inherent characteristics (primarily full width half maxima (FWHM), but also energy dependency and phase delays), the underlying BB noise structure, the overall variability level, and the correlations between them [28]. After examining PDS three main classes of LFQPOs can be found. Firstly, Type-A QPOs (AQPOs), they are identified by a wide, weak peak at ~ 6–8 Hz [28]. Typically, AQPOs emerge in the HSS shortly after the HIS transition. Secondly, a rather strong and narrow peak, with centroid frequencies (CF) of ~ 6 Hz or ~1-3 Hz, is the defining feature of Type-B QPOs (BQPOs) [33]. The SIS is defined basically by the existence of this class and at frequencies 0.1–15 Hz, Type-C QPOs (CQPOs) are distinguished by a sharp, narrow, and fluctuating peak [28]. They are often seen at the luminous end of the LHS or the HIS in addition to the HSS for some accretor. Investigating the AF surrounding BHs and NSs may be done indirectly through the research and analysis of LFQPOs. Several research groups have attempted to use their models, which are based on geometrical effects and instabilities, to explain the formation of LFQPOs. According to transition layer model put forth by Titarchuk and Fiorito [34], CQPOs are caused due to magneto-acoustic (MA) oscillations having viscosity in a spherically bounded transition layer. This layer is created when the material from the disc adjusts to the sub-Keplerian boundary conditions close to the CO. When MA waves propagate within the coronae, causing oscillations, Cabanac et al. [35] explained concurrently CQPOs and the related BB noise using a single model. The accretion-ejection instability (AEI) has been proposed by Varnière, Rodriguez and Tagger [36, 37] to produce A, B, and C QPOs in three distinct regimes: non-relativistic (CQPOs), relativistic (AQPOs, where coexistence of AEI and Rossby wave instability can be observed), and during the transition between the two regimes (BQPOs) [28]. To describe the formation and behaviour of two kHz QPOs and a type of LFQPO (oscillation of horizontal branches) in NS XBs, Stella and Vietri [38] presented a relativistic precession (RP) model. To explain CQPOs and the noise they produce, Ingram, Done and Fragile [39] put up a model based on the RP predicted by the general relativity theory.

- High-Frequency QPO (HFQPO): The first HFQPO in a BH binary is of ~ 67 Hz which has been discovered through the first observation of GRS 1915+105 [40]. HFQPOs only appear in observations with a high accretion rate. A selection effect accounts for at least some of the fact that not every observation with a high accretion rate yields the detection of an HFQPO [28]. Either a single peak or a pair can be observed in them. For examples, GRS J1655-40 shows two distinct peaks but others show only one peak but sometimes some sources like H 1743-322 shows one clear peak with one weak peak [28]. For HFQPOs, the typical fractional root-mean-squared variability is 0.56–0.6%, which increases sharply with energy but for GRS 1915 + 105, it is observed as more than 19% at 20-40 keV energy. The ratio of CF to FWHM of QPO peaks for HFQPO or Q-factor ranges between 5-30 [40]. There are various models that have been proposed for these HFQPO explanations. First the relativistic precession model which was proposed for LFQPOs can explain HFQPOs up to some extent. A family of resonance models was developed by research groups for the study of three sources: GRO J1655-40, XTE J1550-564, and XTE J1743-322 [28].

Understanding QPOs is crucial for unravelling the underlying physical processes. Their relationship with specific spectral state and state transition provides valuable insights. There have been several recent works on the theoretical advancement of QPOs. According to Mondal [41], due to turbulence pressure in the post-shock region, the parameter space expands (in the low angular momentum and low-energy regions) for producing shocks. It has been found by Singh et al. [42] that QPOs generated by oscillating standing shock, forming far from black hole exhibit a weak dependence on BH masses and spins. Additionally, Dihingia et al. [43] found QPOs in the AF of a geometrically thin AD surrounding a Kerr BH (from outside matter is injected) using very long relativistic magnetohydrodynamic simulations. With their simple unified AF model, they were able to detect both LFQPOs (about 10 Hz) and HFQPOs (approximately 200 Hz), as well as a clear distinction in frequency of QPOs throughout the course of the evolution.

V. OBSERVATIONAL TECHNIQUES AND INSTRUMENTS

Due to the enormous energy of X-rays and the severe conditions surrounding accretors, studying X-ray emission necessitates multifaceted observational techniques and instruments. Space-based X-ray observatories are intended to detect X-rays emanating from astrophysical sources. To prevent absorption and scattering they are operated above earth's atmosphere. For pulsating sources, temporal study of X-ray pulses can offer insights on the accretion dynamics and geometry around the central object where we need high resolution X-ray spectroscopy which may reveal the temperature and ionization state of the accreting material, and its surroundings followed by high-resolution imaging. Apart from spectroscopy, X-ray polarimetry is an ongoing field that determines the polarization of X-rays providing information on the geometry of the emitting area and the MFs surrounding the accretor [44]. There are different instruments based on these types of techniques. Notable examples are Chandra (which enables high-resolution imaging and spectroscopy with spatial resolution of 1 arcsec [1]), XMM-Newton (which offers medium to high resolution spectroscopy with spatial resolution of 5 arcsec [1]), NuSTAR (Nuclear Spectroscopic Telescope Array) (which prioritizes on 3-79 keV band with good sensitivity [45]), AstroSat (which enables low to moderate resolution spectroscopy throughout the 0.3-80 keV energy range for X-rays [46]), XRISM (which prioritizes on 0.3-12 keV band with lower resolution than Chandra [47]), ATHENA (which offers good spectral resolution in 0.2-12 keV band [48]).

Table II. Table for comparing features of different instruments.

| Instrument | Energy (keV) | Resolution | Applications | Reference |
|---|---|---|---|---|
| Chandra | 0.1-10.0 | <1 arcsec (Spatial) | Study of high-energy X-rays coming from disc | [49] |
| XMM-Newton | 0.1-15.0 | ~5 arcsec (angular) | Study of accretion mechanism in supergiant XBs. | [50, 51] |
| NuSTAR | 3.0-79.0 | ~18 arcsec (angular) | Survey of BHs and probing the inner part of the accretion disc and constraining BH spins via reflection components. | [52, 53] |
| AstroSat | 0.3-80.0 | 10μs for LAXPC, 2.37 sec (Photon counting mode) and 0.278 sec (fast windowed mode) for SXT (temporal) | Finding of novel transient sources in the X-ray sky, conducting HXR surveys, and gathering BB spectroscopic data on XBs, AGN, SNe remnants, and stellar coronae. | [54, 55, 56] |
| XRISM | 0.3-13.0 | ≤1.7 arcmin (angular) | Study of the formation and evolution of COs, including ultra-luminous sources and magnetic CV. | [57, 58] |
| ATHENA | 0.2-12.0 | 2.5 eV (up to 7 eV) (spectral) | Study of the nature of the primary source of high energy radiation in stellar mass and supermassive accreting BHs (AGN). | [59, 60] |

## VI. FUTURE PROSPECTS AND CHALLENGES

Extreme MFs and gravitational fields are produced by accretors such as NSs and BHs. The characteristics of these scenarios, such as the strength of MFs and the dynamics of acceleration of particles, may be seen through X-ray emissions. We can investigate stars' evolution mechanisms in binary systems that have accretors. The mass of the partner star and the accretor, as well as the influence of mass transfer on how they evolve, may be ascertained by X-ray studies. X-ray research on distant accretors may reveal information about the galaxies' formation and the evolutionary phases of supermassive BH. A short period may result in substantial fluctuations in X-ray emissions from accretors, which makes observation and data analysis complex. Advanced modelling and ongoing observation are necessary to comprehend these distinctions. Next challenge is related to background noise as other cosmic source may contribute noise in the target observation. It is crucial to develop X-ray detectors that are robust and capable of functioning throughout a wide energy band. It is still a technological challenge to improve the spatial and spectral resolution. Machine learning (ML) helps in promising developments in areas of high-energy observation and high-resolution spectroscopy of X-ray emission. To extract astrophysical information of astrophysical sources (from catalogues) requires the automatic classification of X-ray detections. The absence of optical counterparts and representative training sets makes classification in X-ray astronomy still difficult. In order to provide Chandra Source Catalogue (CSC) sources with a limited number of labelled sources with probabilistic classes and without additional information from optical and infrared catalogues, Pérez-Díaz et al. [61] developed a different methodology using an unsupervised ML approach. They presented a probabilistic class catalogue for 8756 sources, encompassing 14507 detections, and showed that the method is successful in detecting emission from young stellar objects and efficiently differentiating between large-scale and small-scale compact accretors. In time-domain astronomy, X-ray observations are essential. A significant factor in time-domain astronomy as well as high-energy astrophysics is the newly launched X-ray astronomical satellite known as the Einstein Probe (EP). Using data from the EP-WXT Pathfinder—Lobster Eye Imager for Astronomy (LEIA) and EP-WXT simulations, Zuo et al. [62] created a ML classifier for independent source classification.

Radiation is released by accreting objects throughout the electromagnetic spectrum [63]. Multi-wavelength observations are therefore crucial for studies of accretors. The inner hot region in the AD emits X-rays, whereas the cooler outer regions emit in the infrared and optical bands, which allow analysis of the entire disc outside the X-ray emitting inner regions [64]. JWST is perfectly suited to investigate the colder outer parts of ADs because of its exceptional sensitivity to infrared light [65, 66]. Researchers can obtain a more thorough knowledge of accretion processes and emission spectrum by combining X-ray observations (using X-ray Telescopes like Chandra, XRISM, ATHENA) with JWST's infrared capabilities.

## VII. CONCLUSION

In this study, we reviewed several kinds of accretors, especially binary systems, investigating their origins, underlying processes, and unique characteristics, including energy spectra and temporal variability. The X-ray viewpoint provides a unique insight into the evolution and correlations of these systems with their surroundings and is an essential tool in decoding the complex processes that take place within them. New discoveries that have the potential to drastically alter our perception of the cosmos will arise from the interplay of technology and astrophysics as we improve techniques for X-ray emission observation from accretors. Promising developments are anticipated in the areas of high energy observations and high-resolution spectroscopy, as well as ML for data processing. These theoretical discussions will be beneficial in understanding relativistic jet features, such as formation of jets and acceleration processes, which may all be investigated in future observations [67]. Strong MFs' role in stabilizing extremely bright thin discs and the behavior of various MF topologies for various accretors are still unknown [68]. However, the scientific community will need to focus its efforts due to challenges with instruments sensitivity, the intricacies of astrophysical processes, and environmental perturbations. This study offers a foundation for future research in this topic by synthesizing the existing level of knowledge on relativistic accretors and high-energy X-ray observations.

ACKNOWLEDGMENT

The authors AJB, LD and BS acknowledge the reviewer for thoroughly reviewing the manuscript and providing beneficial comments. The authors also acknowledge the use of Meta AI and Google Gemini AI to improve the readability of the text.